\documentclass{emulateapj}

\begin{document}

\title{Parameter Estimation and Confidence Regions in the Method of Light Curve
Simulations for the Analysis of Power Density Spectra}

\author{Martin Mueller\altaffilmark{1} and Greg Madejski\altaffilmark{1}}
\altaffiltext{1}{Kavli Institute for Particle Astrophysics and Cosmology,\\ SLAC National Accelerator Laboratory, 
2575 Sand Hill Road, Menlo Park, CA 94025, USA}

\begin{abstract}

The Method of Light Curve Simulations is a tool that has been applied to X-ray monitoring observations of Active 
Galactic Nuclei (AGN) for the characterization of the Power Density Spectrum (PDS) of temporal variability 
and measurement 
of associated break frequencies (which appear to be an important diagnostic for the mass of the black hole in 
these systems as well as their accretion state). 
It relies on a model for the PDS that is fit to the 
observed data. 
The determination of 
confidence regions on the fitted model parameters is of particular importance, and we 
show how the Neyman construction based on distributions of estimates may be implemented in the context of light curve simulations. 
We believe 
that this procedure offers 
advantages over the method used in earlier reports on PDS model fits, 
not least with respect to the correspondence between the size of the confidence region and 
the precision with which the data constrain the values of the model parameters. 
We plan to apply the new procedure to existing \emph{RXTE} and \emph{XMM} observations 
of Seyfert I galaxies as well as \emph{RXTE} observations of the Seyfert II galaxy NGC 4945.

\end{abstract}


\section{INTRODUCTION}

The study of the temporal variability of the 
X-ray flux from accreting black holes has revealed a complex behavior 
of the accretion flow 
\citep[e.g.][]{mushotzky93,remillard06}. 
One widely applied tool for characterizing the variability 
is the Power Density Spectrum (PDS). 
The shape of the broad-band PDS as 
well as the location of identifiable features such as breaks and 
quasi-periodic oscillations provide the observational constraints 
for physical models of the system that generates the variability. 
Of particular interest is the apparent linear scaling between the 
high-frequency break timescale and the black hole mass in these 
systems over many orders of magnitude 
\citep{mchardy04}.

The analysis of the PDS of Active Galactic Nuclei (AGN) is complicated 
by the question of how to 
assign uncertainties to the Fourier amplitudes measured from 
one observed realization of what is presumed to be a stochastic 
process 
\citep{lawrence87,mchardy87}. 
A measure of the expected spread in the observed values is essential 
for the correct interpretation of the Fourier spectrum. 
In addition, intrinsic properties of the Fourier transform (red noise 
leak and aliasing), uneven sampling of the time series, 
and measurement uncertainty in the count rate distort the 
spectrum; these effects need to be corrected for when quantifying the 
shape of the broad-band spectrum. 
A method based on Monte Carlo simulations to determine a reliable 
measure of the PDS uncertainties and to account for these distortions 
was first proposed by 
\citet{done92}. 
The main feature of the method is the concept of simulating the 
possible range of realizations of the underlying process and incorporating 
the shape-distorting effects of uneven sampling, 
red noise leak, and aliasing. Uncertainties on the Fourier spectrum 
are determined using light curves generated from a model for the PDS, 
and the level of agreement between the model and the data is 
quantified by a $\chi^2$ fit statistic. By analogy to X-ray spectral 
fitting, the application of these observational effects to the chosen model 
results in the ``folded model'', which is then used in the comparison to 
the observations.

Subsequent work by 
\citet{uttley02} (hereafter UMP02), 
incorporating the recommendations for the simulation of stochastic 
processes in 
\citet{timmer95}, 
led to a canonical method for the analysis of AGN X-ray light curves, 
obtained mainly from 
\emph{RXTE} and 
\emph{XMM-Newton}. 
The process to be modeled is expressed in the form of a 
parametric expression for 
the PDS; depending on the complexity of the model, a 
varying number of adjustable 
parameters determine the shape and normalization of the model PDS. 
In addition to 
the updated Monte Carlo simulations, the authors present detailed procedures 
for the statistical evaluation of the model fit, i.e. the assignment of a 
goodness-of-fit measure and the derivation of confidence regions on 
model parameters. The fit statistic, which was dubbed the ``rejection 
probability'' by subsequent authors, is now different from a standard $\chi^2$ 
statistic. This development toward a statistically more sophisticated technique 
was influenced by considerations about the resolution of the PDS, which often 
needs to be compromised to satisfy the conditions under which the 
$\chi^2$ statistic may be used safely 
\citep{papadakis93}. 
This new method has found widespread 
applicability; the results reported in 
\citet{markowitz03} (hereafter M03), 
\citet{mchardy04}, 
\citet{markowitz05}, 
\citet{mchardy05}, 
\citet{uttley05b}, 
\citet{mchardy07}, 
\citet{summons07}, 
\citet{arevalo08}, 
and 
\citet{marshall08} 
are all based on it. 
Our initial report on the PDS of NGC 4945 
\citep{mueller04}
similarly took the published method and introduced some additional 
changes.
(In contrast to the above papers, 
\citet{green99}, 
\citet{vaughan03a}, 
\citet{vaughan03b}, 
and
\citet{awaki05} 
implement Monte Carlo simulations for the derivation of uncertainties 
on the PDS, but use the 
standard $\chi^2$ statistic for the model fit.) 

In the general case of fitting a model to a set of data, the derivation of best fit values 
of the model parameters (called ``point estimation'' in statistics) involves the 
identification of the location in parameter space at which the fit statistic attains 
an extremum\footnote{The use of Monte Carlo simulations for the derivation of the 
folded model in the case of PDS fitting results in a fit statistic that can not be expressed in closed form as 
a function of the parameters. 
Numerical methods therefore need to be employed 
to search for the location of the extremum.}. 
These best fit values are called ``estimates''; the recipe for finding 
an estimate for a particular parameter is called the ``estimator''. 
Point estimates by themselves are of limited use. Instead, confidence 
regions on the fitted model parameters 
characterize how well the data constrain the model, and 
goodness-of-fit tests may be applied to test whether the chosen model 
is an adequate description of the data and whether certain models are 
favored over others. 

The definition of a confidence region is crucial to its proper interpretation. 
A confidence region (with associated confidence level $C$) is a region in parameter space computed from the measured 
data that has a probability $C$ of containing the true set of parameter values. 
In other words, if the measurement were repeated, a different confidence region 
would be obtained for each data set, 
but a fraction 
$C$ of them would enclose the (unknown) point in parameter space on which the 
measured data are based. 
This of course assumes that the model under consideration is the correct one; if 
a goodness-of-fit test indicates that the chosen model is a bad description for 
the data, then confidence regions on its parameters are of little value.

Confidence regions are often interpreted as expressing the precision with 
which the model parameters may be determined given the data. 
In practice, for any given fitting procedure, there are usually several plausible methods that produce 
regions with the 
required property to make them confidence regions. A useful additional consideration 
is therefore whether the 
size of the region that a chosen method returns depends appropriately on the measurement 
uncertainties. Furthermore, the value of the fit statistic at the location 
of the best fit should have no or only a weak influence on the 
size.\footnote{By way of example, in a standard $\chi^2$ fit, 
under the assumption that the chosen model is the correct one and that there are no 
systematic errors in the measurement, the minimum value of $\chi^2$ in a given fit 
is fully determined by the ratio of the actual amount of statistical fluctuations in the data 
to the expected amount and does therefore not depend on the 
size of the measurement uncertainties. In situations encountered in practice, 
the conditions under which this is true are often violated to a certain degree, such that 
a small influence on the size of the confidence region cannot be ruled out.}

We review some of the concepts of model fitting using the 
$\chi^2$ statistic in Section 
\ref{sec:chisqu}, 
and we show how, in a simple toy model set-up, the $\Delta \chi^2$ prescription for 
finding confidence regions 
satisfies the consideration above. 
The ``rejection probability'' as a fit statistic is evaluated on the same 
criteria in Section 
\ref{sec:rejprob}, 
and the strong dependence of the size of the confidence region on the minimum 
rejection probability demonstrated. 
We then introduce the Neyman construction based on simulated distributions of estimates 
in Section 
\ref{sec:estimdistr} 
as an alternative to the use of the rejection probability. 
This paper does not present any actual results obtained from the proposed 
method. However, we outline in Section 
\ref{sec:discussion} 
possible changes that may occur if 
PDS model fits obtained using contours of constant rejection probability,
including our own work on NGC 4945, 
were re-examined using the Neyman construction. 
Section 
\ref{sec:conclusion} 
summarizes the paper.

\section{POINT ESTIMATION AND CONFIDENCE REGIONS USING THE $\chi^2$ STATISTIC}
\label{sec:chisqu}

The most familiar fit statistic in Astrophysics is without a doubt the $\chi^2$ statistic. 
It applies well to problems where the measured quantities are Gaussian distributed 
with known uncertainties. Even in cases where that condition is not satisfied, 
the $\chi^2$ statistic can sometimes still yield useful parameter estimates. 
However, its main attraction lies in the ease with which confidence regions on 
fitted parameters can be derived if the distributions are Gaussians, namely through 
the concept of $\Delta \chi^2$ 
\citep[see e.g.][]{lampton76,press,bevington}. 
Any desired significance level $0<\alpha<1$ maps onto a value of $\Delta \chi^2$ 
such that, after determining the best fit values of the model parameters by 
minimizing $\chi^2$, 
the region in parameter space bounded by the surface of constant 
$\Delta \chi^2$ contains the true set of parameter values with a confidence 
$C=1-\alpha$.

Let us illustrate the $\Delta \chi^2$ procedure on a toy model setup
to introduce additional notation that we will refer back to in subsequent sections.

Let $y$ be a physical variable that is expected to be proportional to a single 
independent variable $x$. As part of an experiment, $y$ is measured for a fixed set 
of non-equal $x_i$ ($i=1,...,N$). The measurement is expected to result in 
Gaussian uncertainties on $y$, with a constant standard deviation $\sigma$ independent 
of $i$. Let $\{y_i\}$ 
($i=1,...,N$) be the set of measurements at the corresponding values $x_i$.

We now wish to fit these data with a model $y = k \, x$. The $\chi^2$ fit statistic 
is then a function of the one model parameter $k$ and the set of observed values 
$\{y_i\}$ (all sums are over i from 1 to N):

\begin{equation}
\chi^2(k, \{y_i\}) = \sum \frac{(y_i - k x_i)^2}{\sigma^2}.
\label{eq:chisqu_vanilla}
\end{equation}

Minimizing $\chi^2(k, \{y_i\})$ with respect to $k$ yields the estimate $\hat{k}(\{y_i\})$ and 
the minimum value of the fit statistic $\chi^2_{\rm min}(\{y_i\})$:

\begin{equation}
\hat{k}(\{y_i\}) = \frac{\sum x_i \, y_i}{\sum x_i^2}
\label{eq:k_estimator}
\end{equation}

and

\begin{equation}
\chi^2_{\rm min}(\{y_i\}) = \sum \frac{(y_i - \hat{k} x_i)^2}{\sigma^2}.
\label{eq:min_chisqu}
\end{equation}

Using these two equations, the expression for the change in $\chi^2$ as $k$ is varied 
evaluates to

\newpage
{\setlength\arraycolsep{2pt}
\begin{eqnarray}
\Delta \chi^2(k, \{y_i\}) & \equiv & \chi^2(k, \{y_i\}) - \chi^2_{\rm min}(\{y_i\}) \nonumber\\
& = & \frac{\sum x_i^2}{\sigma^2} \, (k - \hat{k}(\{y_i\}))^2.
\end{eqnarray}

To derive the 68\% confidence region (i.e. the ``1$\sigma$'' uncertainties)\footnote{The term ``1$\sigma$'' is 
sometimes used to indicate the 68\% confidence region even if the fit statistic is not $\chi^2$. In such 
cases, the standard deviation of the distribution of the parameter may not have the same interpretation, 
but the confidence regions parameterized by the confidence $C$ are always well-defined.}
around $\hat{k}(\{y_i\})$ (significance $\alpha=0.32$), we set 
$\Delta \chi^2(k, \{y_i\}) = 1.00$. The resulting region satisfies 

\begin{equation}
|k - \hat{k}(\{y_i\})| < \frac{\sigma}{\sqrt{\sum x_i^2}}.
\label{eq:confreg}
\end{equation}

Since we assumed that the measured $y_i$ are in fact well-described by the model, 
then each $y_i$ has to be drawn from a Gaussian distribution around the true value, i.e. 

\begin{equation}
y_i \sim g(k_{\rm true}\, x_i, \sigma),
\label{eq:randomised_y_i}
\end{equation}

where $g(a, b)$ is a Gaussian distribution with average $a$ and standard 
deviation $b$, and $\sim$ denotes ``drawn from.'' The estimate $\hat{k}(\{y_i\})$ is then drawn from the following 
probability density function: 

\begin{equation}
\hat{k}(\{y_i\}) \sim g\left(k_{\rm true}, \frac{\sigma}{\sqrt{\sum x_i^2}}\right).
\label{eq:estimdistr}
\end{equation}

In other words, if the act of measuring the set $\{y_i\}$ is repeated many times, $\hat{k}(\{y_i\})$ will differ from 
$k_{\rm true}$ by less than $\sigma/\sqrt{\sum x_i^2}$ 
68\% of the time. It should now be immediately obvious that the size of the confidence region 
(Equation 
\ref{eq:confreg}) 
is such that in precisely those 68\% of cases the confidence region includes $k_{\rm true}$, 
confirming what the confidence region was designed to express about the experiment.

We have thus confirmed that in this simple setup, the $\Delta \chi^2$ prescription produces 
intervals for the model parameter $k$ that satisfy the requirements of confidence intervals. Furthermore, 
it can be seen from Equation
\ref{eq:confreg} 
that the size of the confidence interval is proportional to the measurement uncertainty $\sigma$ and 
independent of $\chi^2_{\rm min}$. We defer to existing publications 
\citep[specifically][]{lampton76} 
for the extension of these results to higher-dimensional parameter spaces. We only wish to note here that the 
independence of the size of the 
confidence region on $\chi^2_{\rm min}$ is guaranteed through the independence of the distribution of 
$\Delta \chi^2$ on $\chi^2_{\rm min}$, as demonstrated in 
\citet[Appendix IV]{lampton76}.

\section{THE REJECTION PROBABILITY}
\label{sec:rejprob}

The measurement of the level of agreement between the model and the observed data 
in the method of light curve simulations in UMP02 
relies on a statistic called by subsequent authors the ``rejection probability.'' 
It is defined analogous to a $p$-value, with the rejection probability being 
one minus the $p$-value of the measured $\chi^2_{\rm dist}$ fit statistic. 
Differences in best-fit rejection probability between different models are used 
to favor one model over the others (e.g. a broken power law model compared to 
an unbroken power law model), and 
regions in parameter space where the rejection probability is less than a certain 
value (e.g. 90\%) are then taken as the confidence regions for the fitted model 
parameters.

\subsection{Confidence Regions from Rejection Probability}
\label{sec:rejprob_confreg}

By analogy to $\chi^2$ fitting, the UMP02 
procedure for determining confidence regions is equivalent to 
identifying the region in parameter space where $\chi^2$ is less than some 
critical value. For the $c$\% confidence region, this critical value is 
simply the $c$-th percentile of the $\chi^2$ distribution with the 
appropriate number of degrees of freedom and is thus 
\emph{independent of the minimum value of $\chi^2$ obtained in the fit.} 
The reason why the authors do not rely on percentiles of $\chi^2$ distributions 
for the determination of confidence regions 
is that the effective number of degrees of freedom varies with position in 
parameter space. Deciding on the basis of 
$p$-values whether a certain point in parameter space is included in the 
confidence region is therefore more robust.

The region produced in this manner do have the required property to make them 
confidence regions, i.e., they include the true value of the parameters with the 
desired probability. 
However, their sizes depend strongly 
on the value of the fit statistic at the location of the best fit. If the minimum 
rejection probability 
in a fit is just below 90\%, 
the contours of 90\% rejection probability will 
be found fairly close around the best fit, leading to the erroneous conclusion that 
the data lend themselves to the placement of very precise limits on the model 
parameters. Note that a 
minimum rejection probability of 90\% does not by itself indicate a bad fit, since 
there is still a 10\% chance of obtaining a fit as bad or worse due simply to 
statistical fluctuations; we \emph{are} therefore justified in searching for the 
confidence region associated with the parameters of such a fit. Conversely, 
if the minimum rejection 
probability is very low, the 90\% contours will enclose a large area. Furthermore, if the 
minimum rejection probability is above 90\%, there will be no 90\% confidence 
region at all.

\begin{figure}
\plotone{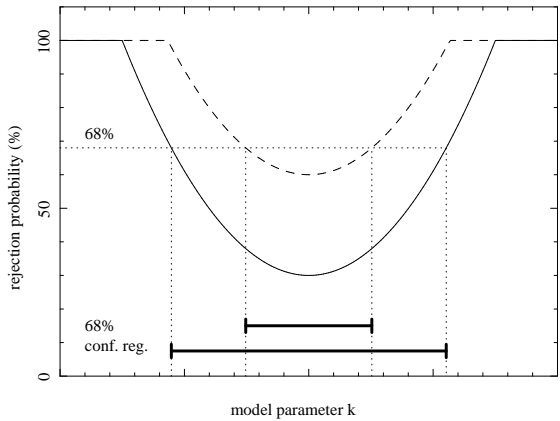}
\caption{Schematic plot illustrating the dependence of the size of the confidence region on the minimum value of the rejection probability. 
The solid and dashed lines are stylized representations of the behavior of the rejection probability as a function of a model parameter $k$ 
for fits to two different data sets. Both fits return the same estimate for $k$, but due to statistical fluctuations, the fit indicated 
by the solid line results in a significantly lower minimum rejection probability. The procedure for determining the 68\% confidence interval on $k$ 
for each fit is indicated by the dotted lines, showing the projection of the intersection points between the line of 68\% rejection probability 
and the respective parabola onto the $k$ axis. Even though this is only a schematic representation, and the detailed behavior of the rejection 
probability as a function of any of the parameters in a real fit may be more complicated, the negative correlation between the size of the 
confidence region and the minimum value of the rejection probability is expected in general.
\label{fig:rejprob}}
\end{figure}

We illustrate the inverse correlation between the minimum rejection probability and the 
size of the resulting confidence region schematically in Figure 
\ref{fig:rejprob}. 
This behavior is apparent in some of the published results 
(UMP02, M03, 
\citealp{mchardy07}, 
\citealp{summons07}).

\subsection{The Empirical Distribution of $\chi^2_{\rm dist}$}
\label{sec:rejprob_gof}

The calculation of the rejection probability 
relies on an approximate determination of the empirical distribution of $\chi^2_{\rm dist}$: 
Since the effective number of degrees of freedom is a function of the model parameters, 
the distribution is rightfully calculated separately for 
each grid point in parameter space. 
However, the $\chi^2_{\rm dist}$ values for the 
simulated Fourier spectra are only calculated at their original 
grid point and are \emph{not} the best fit values found by performing the point 
estimation on each simulated spectrum 
\citep[see e.g.][Section 15.6]{press}.

The approximation was most likely introduced due to considerations about computing time, 
because the minimization of $\chi^2_{\rm dist}$ over the parameter space for the hundreds of simulated 
spectra that are typically involved incurs a significant
computational load. However, given the advances in computer power since the original method 
was introduced, the reason for the approximation may well have fallen away.

\section{CONFIDENCE REGIONS AND GOODNESS-OF-FIT TEST USING SIMULATED DISTRIBUTIONS OF ESTIMATES}
\label{sec:estimdistr}

We propose the following set of procedures, most importantly the Neyman construction based on 
distributions of estimates for 
the derivation of confidence regions, 
as an alternative to the use 
of the rejection probability for PDS model fits. The new method 
returns confidence 
intervals whose size has the desired property of being independent of the value of the fit 
statistic at the location of the best fit. Furthermore, it deals very naturally with biased 
estimators\footnote{Biased estimators are estimators with 
an expectation 
value different from the true one. The $\hat{k}$ estimator used in Section 
\ref{sec:chisqu} is unbiased because its expectation value is $k_{\rm true}$ (Equation 
\ref{eq:estimdistr}). In more complicated situations, such as the PDS fits under consideration here, 
one does not generally know a priori whether the chosen estimators are biased or not.}.

Throughout this section, it is assumed that the 
$\chi^2_{\rm dist}$ fit statistic can be calculated 
for an arbitrary point in parameter space through 
the use of simulated light curves. The procedures 
are however not specific to the $\chi^2_{\rm dist}$ fit statistic; any other 
statistic which attains an extremum at the location of the best fit may 
be substituted for $\chi^2_{\rm dist}$.

\subsection{Point Estimation}
\label{sec:point_estimation}

$\chi^2_{\rm dist}$ may be used directly for point estimation, i.e. the 
estimates $\hat{\Theta}_{\rm obs}$ for the parameters of the model used to 
describe the observed Fourier spectrum $P_{\rm obs}(\nu)$ 
are the values 
of the parameters at the grid point that minimize $\chi^2_{\rm dist}$. 
The estimates $\hat{\Theta}_{\rm sim}$ for any of the simulated light curves 
(used further below) 
can be found similarly by substituting the simulated spectrum in place of 
the observed spectrum and minimizing $\chi^2_{\rm dist}$ over the parameter 
space.

\subsection{Goodness-of-Fit and Hypothesis Testing}
\label{sec:gof}

In order to test whether the minimum $\chi^2_{\rm dist}$ value of the observed Fourier 
spectrum signifies an acceptable fit, we 
use the simulations to determine the distribution from which $\chi^2_{\rm dist}$ 
is drawn. The null hypothesis is that the measured Fourier spectrum was in fact 
produced by the model under consideration. 
Let $\Theta_{\rm best}$ be our best guess for the true values of the parameters, i.e. the grid point 
closest to the center 
of the confidence region. (If the estimators are unbiased, $\Theta_{\rm best}$ can be set 
equal to $\hat{\Theta}$.) 
For each of the simulated light curves generated for $\Theta_{\rm best}$, 
we record its best fit $\chi^2_{\rm dist}$ (already found above in the determination of the 
confidence region). The goodness-of-fit measure is then 
the familiar $p$-value of the observed spectrum's minimum 
$\chi^2_{\rm dist}$ compared against this distribution of simulated $\chi^2_{\rm dist}$ values. 
As such, it expresses the probability that 
a $\chi^2_{\rm dist}$ value at least as high as the measured one would be obtained by chance; 
a $p$-value smaller than the desired significance level (e.g. 5\%) indicates that the null 
hypothesis can be rejected.

The reason for choosing $\Theta_{\rm best}$ over any other grid point is that the distribution of the 
minimum values of 
$\chi^2_{\rm dist}$ may depend on $\Theta$. In standard $\chi^2$ fitting, 
the distribution of $\chi^2_{\rm min}$ is independent of $\Theta$, being in fact the $\chi^2$ 
distribution with 
the appropriate number of degrees of freedom. In the framework of Fourier spectral fits, this 
independence appears to be broken, such that the effective number of degrees of freedom is a 
function of the model parameters, plausibly because the degree to which adjacent bins in the Fourier 
spectrum are correlated depends on the amount of red noise leak (Mueller et al., \emph{in preparation}). 
Using $\Theta_{\rm best}$ ensures that the 
$\chi^2_{\rm dist}$ distribution thus found approximates as closely as possible the one from which the 
measured $\chi^2_{\rm dist}$ was in fact drawn.

Note that, up to the approximation to the distribution of $\chi^2_{\rm dist}$ used by UMP02, 
this procedure is essentially equivalent to the calculation of the rejection probability. 
The $p$-value is however only used as a goodness-of-fit measure and not for finding the confidence 
intervals.

If more than one model are under consideration to explain the measured data, e.g. when 
one would like to test for the presence of a break in the Fourier spectrum, a decision 
statistic for hypothesis testing needs to be set up. In the framework of the $\chi^2_{\rm dist}$ 
fit statistic, the difference in best fit $\chi^2_{\rm dist}$ values between two models is a 
natural choice for such a statistic (by analogy to the F-test for the $\chi^2$ fit statistic). 
The simulations can once again be used to determine the 
distribution of this difference, from which the critical value corresponding to a desired power 
of the test (``statistical significance'') may be derived. We do not further elaborate on this 
procedure here, since the numbers and decisions involved depend on a balance 
between the sensitivity and 
specificity of the test that can only be calculated using actual simulations.

\subsection{Confidence Regions}
\label{sec:confreg}

\begin{figure*}
\plottwo{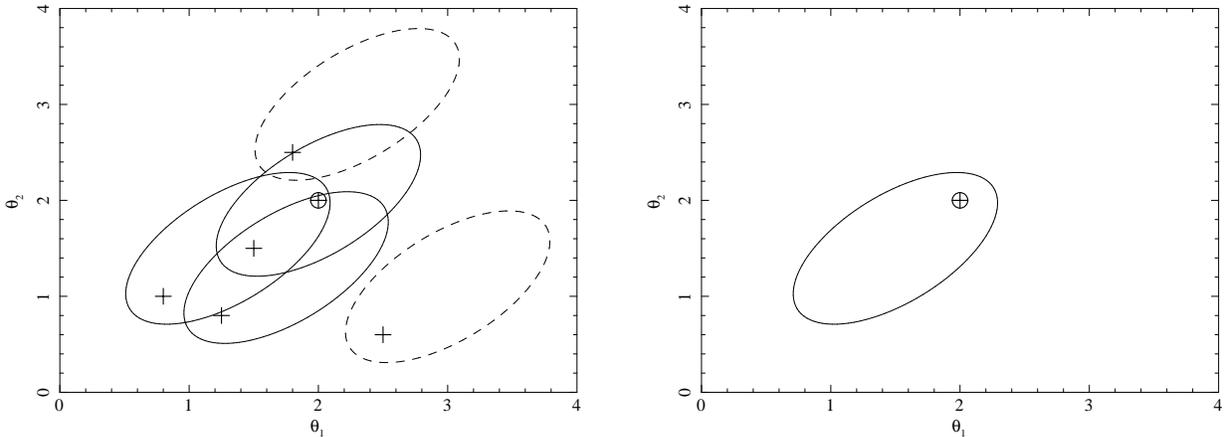}{f2b.eps}
\caption{Schematic illustration of the Neyman construction applied to a model fit with two adjustable parameters $\theta_1$ and $\theta_2$. 
The plot on the left shows elliptical regions obtained from the distribution of the estimates that encompass an unspecified, but constant, 
fraction $C$ of the distribution. The corresponding trial values of the parameters for each ellipse are indicated by the crosses. 
The offset between the crosses and the centers of the ellipses implies a bias in the estimators, kept 
constant as a function of the parameters in this simple example. The location of the observed best-fit 
($\hat{\theta_1}$, $\hat{\theta_2}$) 
is denoted by the cross-hairs. The solid ellipses include the observed best-fit values of the parameters, 
the dashed ones do not. By the prescription of the Neyman construction, the parameter values associated with the solid ellipses are added to the confidence 
region, the others are not. The plot on the right shows the elliptical confidence region (confidence = $C$) that would be obtained if this procedure were to be repeated for 
all possible trial values of the parameters. 
The observed best-fit values are once more 
indicated by the cross-hairs. Note how the estimator bias identified earlier results in a confidence region whose 
center is offset from the observed best-fit values.
\label{fig:neyman}}
\end{figure*}

We implement the Neyman construction 
\citep{neyman37} 
based on simulated distributions of estimates to find confidence limits on model parameters: 
Let $C$ be the desired confidence, e.g. 68\% or 90\%, and $\hat{\Theta}_{\rm obs}$ the 
estimates for the observed Fourier spectrum as found above. Consider now an arbitrary grid point 
in parameter space, $\Theta_{\rm trial}$. Using the simulated light curves generated for that point, 
we can determine the distribution of estimates $\hat{\Theta}_{\rm sim}$ and derive 
a region in parameter space that encloses a fraction $C$ of them. If $\hat{\Theta}_{\rm obs}$ is 
inside that region, $\Theta_{\rm trial}$ is included in the confidence region, otherwise it is not. 
Figure 
\ref{fig:neyman} 
shows this graphically in an imagined two-parameter model fit.

It is easy to see that the size of the confidence regions thus obtained is independent of the value of the 
minimum $\chi^2_{\rm dist}$; no information 
about the measured data enters the calculation of the distribution of estimates, and only the 
estimates $\hat{\Theta}_{\rm obs}$ are used in the subsequent mapping of the confidence region. 
Also, the distribution of estimates, being a measure of the possible range of parameter values that 
might be observed given the design of the measurement, will be broad or narrow depending on the 
uncertainties in the observed Fourier spectrum. Therefore, the size of the confidence region will scale 
with the uncertainties, preserving the intended interpretation that the size of the confidence region 
expresses the precision with which fitted model parameters can be determined.

As a 
consequence of using distributions of estimates in the Neyman construction, biased estimators 
will be corrected, and the region returned by the algorithm has the 
required property of enclosing the true values of the parameters with probability $C$. 
In principle, if any of the estimators are strongly biased, $\hat{\Theta}_{\rm obs}$ 
might lie outside the confidence region. The 
confidence region is however always more meaningful in the determination of 
probable values of model parameters than a (possibly biased) point estimate.

Note also that the shape of the region enclosing the fraction $C$ of all simulations may be 
freely chosen by the user. This freedom is intrinsic to the Neyman construction. 
However, in order to obtain the tightest possible constraints on the 
parameters, the smallest region should be used.

A complication arises out of the limitation of being able to evaluate $\chi^2_{\rm dist}$ 
only on a grid in parameter space. The distributions of the estimates are therefore composed 
of finite-size volume elements centered on each grid point. The use of smoothed approximations 
to the empirically derived distributions can eliminate this step-wise behavior; for 
a one-dimensional parameter space, if the distribution of the estimate is sharply peaked, one might 
use a Gaussian fit and determine from the fitted average and standard 
deviation whether the trial point lies inside the confidence interval. Similarly, in two 
dimensions, the use of ellipses fitted to the empirical distribution to enclose the 
required fraction of simulated estimates might be appropriate.

Let us illustrate the procedure of finding confidence regions using simulated distributions of estimates 
on the toy model introduced in Section 
\ref{sec:chisqu}. 
The difference to the procedure described above is that the estimate $\hat{k}$ can be found 
analytically and does not rely on a grid search using simulations. However, let us suppose that 
simulations were set up for this simple problem. We use Equation 
\ref{eq:k_estimator} 
to calculate $\hat{k}_{\rm obs}$ for the measured data $\{y_i\}$. Let $k_{\rm trial}$ be an arbitrarily chosen 
real number, for which we want to determine whether it is inside the confidence interval. Using 
simulations with $k_{\rm trial}$ as the ``true'' parameter value (i.e. randomizations of the observations 
as given by Equation 
\ref{eq:randomised_y_i}), we would then find (Equation 
\ref{eq:estimdistr})  that the probability density function of the estimates for this trial value is given by
$g(k_{\rm trial}, \sigma/\sqrt{\sum x_i^2})$, 
i.e. a Gaussian distribution centered on $k_{\rm trial}$. (In this toy model, the distribution of the 
estimates is translationally invariant under changes in $k_{\rm trial}$; this feature is not expected in 
general.) 
The smallest interval enclosing 68\% of the simulations is comprised of the values within one standard 
deviation from $k_{\rm trial}$, and by the prescription of the method $k_{\rm trial}$ is included 
in the confidence region if and only if

\begin{equation}
|k_{\rm trial} - \hat{k}_{\rm obs}| < \frac{\sigma}{\sqrt{\sum x_i^2}},
\end{equation}

thus recovering the confidence region in Equation 
\ref{eq:confreg}.

\subsubsection{The Confidence Interval for the Model Normalization}
\label{sec:model_normalization}

The model to be fit to the data usually includes an overall normalization factor that carries 
through to the model prediction as a multiplicative factor. In this situation, the derivation of 
confidence intervals on the model normalization can be simplified. In practice, simulations need only 
be done once for an arbitrary normalization, since the model prediction and uncertainties for any 
other normalization 
may be calculated simply by scaling. (For a discussion of the complications introduced by 
measurement uncertainties, see Appendix 
\ref{sec:norm_and_pl}.) 
For point estimation, the best fit normalization at 
any point in parameter space\footnote{The parameter space now under consideration excludes the 
model normalization as a parameter, due to its special treatment.} can easily be found, either 
analytically or through a numerical search. This procedure is unchanged from UMP02 (section 4.2). 
The estimates for the remaining parameters, either for 
the measured data or for 
any of the simulated spectra that may be substituted for it, are then once again the 
values of the parameters at the grid point in the remaining parameter space where $\chi^2_{\rm dist}$ 
attains a minimum.

For the derivation of confidence regions, we again wish to determine whether a certain grid point 
in parameter space, 
with normalization $N_{\rm trial}$ and remaining parameters $\Theta_{\rm trial}$, is included at 
a given significance level. 
Let $N_0$ be the original normalization with which the light curves at 
$\Theta_{\rm trial}$ were simulated, and $f_0(\hat{N},\hat{\Theta})$ the 
corresponding probability density function 
of the estimate distribution 
with its dependence on the normalization estimate $\hat{N}$ and the estimates of the remaining 
model parameters 
$\hat{\Theta}$. Because of the multiplicative nature of the model 
normalization, this distribution becomes 
scale-invariant along the $\hat{N}$ axis (see Appendix 
\ref{sec:scale_invariance}), 
such that if a different normalization had been used to simulate those light curves, the distribution 
would simply be an appropriately scaled version of $f_0(\hat{N},\hat{\Theta})$.

The point estimation on the observed Fourier spectrum 
defines the location of the best fit, given by ($\hat{N}_{\rm obs}$, $\hat{\Theta}_{\rm obs}$). 
We now determine the region $R$ in the ($\hat{N}$, $\hat{\Theta}$) space that encloses the required fraction $C$ 
of the estimate distribution (e.g. 68\%). 
Now consider the line in parameter space along which $\hat{\Theta} = \hat{\Theta}_{\rm obs}$, i.e. 
the line parallel to the $\hat{N}$ axis that passes through the best fit. This line either intersects the 
boundary of $R$ in a finite number of points, or else no intersection points exist. The region $R$ is in 
most cases convex, such that there are either zero or two intersection points. We will only 
consider these two cases here; the procedure is easily generalized to non-convex regions that may result 
in additional intersection points.

In the case of zero intersection points, 
($N_{\rm trial}$, $\Theta_{\rm trial}$) is excluded from the confidence region for all values of 
$N_{\rm trial}$. In the other case, let us denote the values of the normalization at the two intersection 
points by $\hat{N}_{\rm 0, low}$ and $\hat{N}_{\rm 0, high}$. Because of the scale-invariance of the 
estimate distribution, these values are proportional to the original normalization $N_0$, such that, for 
any other normalization $N$ that could have been used to generate the light curves, 

\begin{equation}
\hat{N}_{\rm low}(N) = \frac{N}{N_0} \, \hat{N}_{\rm 0, low}
\end{equation}

and

\begin{equation}
\hat{N}_{\rm high}(N) = \frac{N}{N_0} \, \hat{N}_{\rm 0, high}.
\end{equation}

The condition for ($N_{\rm trial}$, $\Theta_{\rm trial}$) to be included in the confidence region now 
reduces to whether the observed value of the normalization is located between these two bounds, i.e. 

\begin{equation}
\hat{N}_{\rm low}(N_{\rm trial}) \le \hat{N}_{\rm obs} \le \hat{N}_{\rm high}(N_{\rm trial}),
\end{equation}

which is equivalent to the condition on $N_{\rm trial}$ 

\begin{equation}
N_0 \, \frac{\hat{N}_{\rm obs}}{\hat{N}_{\rm 0, high}} \le N_{\rm trial} \le N_0 \, \frac{\hat{N}_{\rm obs}}{\hat{N}_{\rm 0, low}}.
\label{eq:Ntrial}
\end{equation}

In summary, the estimate distribution found from light curves simulated at $\Theta_{\rm trial}$, with 
normalization $N_0$, may be used to derive the bounds $\hat{N}_{\rm 0, low}$ and $\hat{N}_{\rm 0, high}$, from which the 
limits on $N_{\rm trial}$ (for a given $\Theta_{\rm trial}$) may be calculated. 
The confidence region 
in the full ($N$, $\Theta$) parameter space finally may be mapped out by repeating the procedure for different values of 
$\Theta_{\rm trial}$.

\section{DISCUSSION}
\label{sec:discussion}

Applying the criterion whether the procedure for determining confidence regions returns regions whose size scales 
appropriately with the uncertainties in the data, we believe that the Neyman construction based on simulated distributions 
of estimates offers a viable and advantageous alternative to the use of the rejection probability. 
While neither UMP02 
nor the authors of the subsequent papers 
utilizing the method make specific claims regarding the statistical properties of the 
regions obtained, anyone not familiar with the data analysis at a sufficient level of detail 
will tend to interpret the quoted uncertainties on the best-fit 
values of the model parameters 
as indicative of the precision with which the data constrain those values. 

We wish to stress however that this does 
not imply that previously reported results are inherently flawed. The confidence limits on break frequencies and power law indices 
for AGN PDS fits may turn out to be different under the application of the new method, but it remains to be seen whether any of 
these changes are large enough to substantially change the interpretation of the observations. Specifically, we do not 
expect that the linear scaling between break timescale and black hole mass 
\citep{mchardy04} 
will be affected even if the confidence intervals on some of the data points were to be modified.

It is likely that the precision with which the break frequencies for Fairall 9, NCG 4151 (both from M03), 
and MCG-6-30-15 (UMP02)
have been reported is too optimistic. Similarly, the confidence regions for the peak frequencies of the 
Lorentzians in the fit to the PDS of Ark 564 may be too small, especially considering that even a small increase in the size 
of the confidence contour 
in a plot where the axes are the logarithms of the peak frequencies (Figure 9 in their paper) has a disproportionate effect 
on the uncertainties on the frequencies themselves. 
On the other hand, some break frequencies may in fact be better determined with current data 
than reported in the literature. Examples of fits where the minimum rejection probability turned out to be particularly low 
include NGC 5548 and NGC 3516 (M03).

A related issue is the use of contours of constant rejection probability as confidence limits on combinations of parameters, such 
as the ratio of break frequencies (Figure 
11 in 
\citealp{mchardy05} 
and Figure 6 in 
\citealp{uttley05b}). 
Limits on this ratio are used as key pieces of evidence to motivate the association of the AGNs under consideration (NGC 3227 and 
MCG--6-30-15) with the analogue 
of the high/soft accretion state in galactic black hole X-ray binaries. Given that the confidence regions were derived using the 
rejection probability, the quoted confidence values with which certain ranges of ratios are excluded in those reports may or 
may not in fact be supported by the data. We do not expect that the use of the new method will alter the general direction of 
these results, i.e. that the ratio of break frequencies in these AGN is likely to be higher than expected for the low/hard state, 
but a re-analysis of the observations focusing on the doubly-broken power law model might be warranted. 
The calculation of the statistical significance with which the model where the ratio of these 
break frequencies is fixed at a value of 30 may be rejected in favor of the original model where both break frequencies 
are allowed to vary forms an additional test on these data. If both of these lines of evidence 
produce mutually consistent results, the case for the classification of these AGN as 
analogues of galactic X-ray binaries in the high/soft state will be strengthened.

On the question of the statistical significance of breaks in the PDS of AGN, we believe that additional work is needed 
to make the values that have been reported more secure. 
Table 5 in 
M03 lists the quantity $\Delta \sigma$ that was designed to express the increase in likelihood of the fit once a break is added 
to the PDS model. It is however not clear from the description whether $\Delta \sigma$ was calculated using the rejection 
probability or the underlying $\chi^2_{\rm dist}$ values. As outlined in Section 
\ref{sec:gof}, 
the $\chi^2_{\rm dist}$ fit statistic does lend itself to the formulation of such a hypothesis test. A validation of 
critical values of differences in $\chi^2_{\rm dist}$ and their corresponding statistical significances will be required. This 
includes using the simulated light curves to calculate type I and type II errors (rate of false positives and false negatives) or, equivalently, 
the specificity and sensitivity of the hypothesis test. As far as we are aware, the amount and quality of data needed to effect a reliable 
detection of a break at a significance level of 5\%, say, is an unanswered question. A systematic investigation in this area, using both real and 
simulated data, might uncover general considerations that would be invaluable for the design of future observatories for AGN timing
research.

The Monte Carlo method for calculating folded models to include observational 
effects may have applications outside of PDS fits to 
AGN X-ray light curves. In particular, the associated procedure of finding 
confidence limits on fitted parameters using simulated estimate distribution 
could be used for X-ray or $\gamma$-ray spectral fits in the low counts-per-bin 
limit, where the discrete nature of the Poisson process becomes important. 
The study of the PDS of galactic X-ray binaries may benefit from an 
application of the Monte Carlo method as well. The shorter 
time scales for characteristic variations and the extensive archive of X-ray 
observations allow for a much more detailed investigation into the shape 
of the broad-band variability spectrum, including the direct 
observation of independent realizations of the underlying 
process 
\citep[e.g.][]{pottschmidt03,done05}. 
The measurement of the distribution of power in individual frequency bins 
is of particular interest in this case. 
Competing physical models for 
the variability in these sources predict distinctive properties of the 
stationarity and degree of stochasticity of the process underlying the 
observations
\citep[e.g.][]{poutanen99,maccarone02,minutti04,uttley05a,zycki05}. 
Adopting the Monte Carlo simulations for the analysis of galactic X-ray 
binaries, specifically the comparison between predicted and observed 
distributions of the Fourier amplitude, may lead to 
tests of certain elements of these models. 
Furthermore, tools beyond the PDS for the investigation of these kinds of stochastic processes, 
such as the bispectrum 
\citep{vaughan07}, 
are more sophisticated in their treatment of the 
underlying variability process, but they will also continue to, at least for a while, 
produce results that are 
not as validated in their interpretation as those derived from standard Fourier analysis. 
Monte Carlo simulations will likely remain essential for the important comparison of these 
tools to the PDS; a solid statistical foundation is in turn essential for these simulations.

\section{CONCLUSION}
\label{sec:conclusion}

Evaluated by the criterion whether the sizes of the confidence regions express the 
precision with which the data constrain the model parameters, we have shown that the 
use of simulated distributions of estimates (Section 
\ref{sec:confreg}) 
is preferable to the rejection probability (Section 
\ref{sec:rejprob_confreg}). Confidence regions determined from the latter 
do have the required property of enclosing the true values of the parameters with 
the given probability, but their size is highly variable depending on the minimum 
value of the fit statistic at the location of the best fit. 
The method based in the former is computationally more 
intensive, but is the only way known to us to derive meaningful uncertainties on fitted model 
parameters in the absence of a better-understood fit statistic.

The end products of the application of the set of procedures in 
Section 
\ref{sec:point_estimation} 
through 
\ref{sec:confreg}  
to the observations 
of an AGN are the best fit values of the parameters for the model-dependent description 
of the PDS, the associated 
confidence limits, and the goodness-of-fit of the model ($p$-value of 
the minimum $\chi^2_{\rm dist}$). Depending on the nature of the investigation, 
more sophisticated statistical tests may be employed to test different hypotheses 
against the same data set, or to quantify the observed variations in the 
parameter values between different AGN.

We are in the process of applying the new method to the 
\emph{RXTE} observation of the Seyfert II galaxy NGC 4945 for which we reported first 
results in 
\citet{mueller04} 
and plan to 
re-analyze the existing archival \emph{RXTE} and \emph{XMM-Newton} observations 
of Seyfert I galaxies with the updated procedure. While we don't expect the conclusions 
drawn from the analysis of these observations to change significantly, this 
will put the investigation into the shape of the PDS 
in AGN on a statistically more solid foundation and 
make the interpretation of the results easier.

\newpage
\section*{ACKNOWLEDGMENTS}

We are indebted to the anonymous referee for 
much valued input on an earlier version of this 
publication, especially concerning the development of 
the analysis method based on the rejection 
probability. 
We would furthermore like to acknowledge Alex Markowitz for 
providing the initial impetus for the critical 
examination of the rejection probability, and 
Chris Done and Piotr \.Zycki for the ongoing 
collaboration on the analysis of the NGC 4945 
observations. 
We also thank Jeff Scargle, and especially 
Frank Porter, for many fruitful 
discussions on statistical techniques and 
their expertise on the associated terminology. 
This research was supported in part by the 
U.S. Department of Energy Contract 
DE-AC02-76SF00515 to the 
SLAC National Accelerator Laboratory.

\appendix

\section{THE SCALE-INVARIANCE OF THE ESTIMATE DISTRIBUTION FOR THE MODEL NORMALIZATION}
\label{sec:scale_invariance}

The special treatment of the model normalization in the derivation of confidence regions relies 
on a property of the estimate distribution under the conditions mentioned in the text 
(Section 
\ref{sec:model_normalization}), 
namely that the normalization carries through to the model prediction as a multiplicative factor.

Let $N_0$ be the normalization (hereafter called the ``input normalization'') 
that was used to generate a set of light curves at an 
arbitrary point in parameter space $\Theta_{\rm trial}$, and 
let $\{P_{\rm in}(\nu_i)\}$ be the Fourier spectrum of one of them, where $\nu_i$ are the 
frequencies over which the spectrum is measured. Additionally, let 
($\hat{N}_{\rm in}$, $\hat{\Theta}_{\rm in}$) be the estimates for this light curve that were 
found as part of the procedure to determine the estimate distribution (see Section 
\ref{sec:confreg} 
for details). 
Because the normalization 
is an overall multiplicative factor in the generation of these light curves, 
the $\{P_{\rm in}(\nu_i)\}$ values are proportional to $N_0$. 

The model to be fit to these data can be written as 
$P(\nu, N, \Theta) = N \, P_r(\nu, \Theta)$, where $N$ is the model normalization and 
$P_r(\nu, \Theta)$ the function describing the dependence of the model on the remaining 
parameters $\Theta$. The folded model at the point in parameter space given by 
$N$ and $\Theta$ is summarized in two variables for each frequency bin: the average 
power 
$\overline{P_{\rm sim}}(\nu_i, N, \Theta)$ and the standard deviation 
$\Delta P_{\rm sim}(\nu_i, N, \Theta)$ 
(for details, see Section 4.2 in UMP02). Both of these scale with $N$:

\begin{equation}
\overline{P_{\rm sim}}(\nu_i, N, \Theta) = N \, \overline{P_{{\rm sim}, r}}(\nu_i, \Theta),
\end{equation}

and

\begin{equation}
\Delta P_{\rm sim}(\nu_i, N, \Theta) = N \, \Delta P_{{\rm sim}, r}(\nu_i, \Theta),
\end{equation}

where $\overline{P_{{\rm sim}, r}}(\nu_i, \Theta)$ and $\Delta P_{{\rm sim}, r}(\nu_i, \Theta)$ 
constitute the folded model for $N = 1$. The fit statistic 

\begin{equation}
\chi^2_{\rm dist}(N, \Theta, \{P_{\rm in}(\nu_i)\}) = 
\sum_i \left( \frac{P_{\rm in}(\nu_i) - N \, \overline{P_{{\rm sim}, r}}(\nu_i, \Theta)}
{N \, \Delta P_{{\rm sim}, r}(\nu_i, \Theta)} \right)^2
\end{equation}

is invariant under changes in 
the input normalization $N_0 \rightarrow \eta \, N_0$ ($\eta > 0$) if the same multiplicative factor 
is applied to the model normalization $N$.
As a consequence, since $\hat{N}_{\rm in}$ and $\hat{\Theta}_{\rm in}$ 
are the estimates for this simulated light curve for $\eta = 1$, then ($\eta \, \hat{N}_{\rm in}$) 
and $\hat{\Theta}_{\rm in}$ would have been 
the estimates if the original normalization had been different by a factor $\eta$. 
This applies to all simulated spectra; therefore the 
distribution of the estimates ($\hat{N}_{\rm in}$, $\hat{\Theta}_{\rm in}$) 
will be scale-invariant along the 
$\hat{N}$ axis: Let $f_0(\hat{N}, \hat{\Theta})$ be the estimate distribution for the original input 
normalization $N_0$ (i.e. $\eta = 1$). For any other value of $\eta$, the 
estimate distribution is then

\begin{equation}
f(\hat{N}, \hat{\Theta}) = \frac{1}{\eta} \, f_0\left(\frac{\hat{N}}{\eta}, \hat{\Theta}\right).
\end{equation}

Note that the above does not require that the normalization be uncorrelated with the other 
model parameters. The invariance of the fit statistic 
is preserved even if such correlations exist.

\subsection{Influence of Measurement Uncertainties}
\label{sec:norm_and_pl}

In the context of PDS model fits, the 
measurement uncertainties 
in the light curve manifest themselves as an additional noise component 
in the Fourier spectrum (Poisson 
level). The scaling of the model prediction with the normalization 
factor is only approximate 
in this case, since the Poisson level is constant and does not scale 
with the model normalization. 
However, the 
intrinsic variability in the light curve by design usually dominates 
over the Poisson level. 
The confidence interval on the model normalization derived while 
ignoring this complication 
is therefore expected to approximate closely the more correct one that 
would be obtained through the 
usual prescription of simulating light curves with different 
normalizations and deriving the 
distribution of the estimates in each case.

\subsection{Applicability to PDS with Logarithmically Averaged Power}

In the canonical method of UMP02, 
$\overline{P_{\rm sim}}(\nu_i)$ is actually the average of the logarithm of the periodogram power, which 
is motivated by the considerations in 
\citet{papadakis93}. 
The logarithm of the model normalization $N$ therefore enters the model prediction as an additive 
constant, while 
the uncertainties $\Delta P_{\rm sim}(\nu_i)$, being standard deviations on what are now logarithmic power 
values whose spread is unaffected by the model normalization, are independent 
of $N$. 
The estimate distribution is then \emph{translationally
invariant} along the $\log \hat{N}$ axis, and the expression for the bounds on $N_{\rm trial}$ 
turns out to be the same as for the linear case 
(Equation 
\ref{eq:Ntrial}).

Note however that 
different \emph{numerical} values for these bounds may be obtained depending on whether the estimate 
distribution is 
expressed as a function of $\hat{N}$ or $\log \hat{N}$. The shape and extent of the region $R$ 
encompassing the desired 
fraction of the estimate distribution may vary; the smallest such region for example will in general be 
different depending 
on the choice of variables. In a complete description of the analysis method, it will be important to state 
which variable 
was used.

\end{document}